\documentclass[12pt,preprint]{aastex}
\begin{document}
\received{}
\accepted{}
\revised{}
\title{A Double-Mode RR Lyrae Star with a Strong Fundamental Mode
Component}
\author{Lindsay Oaster and Horace A. Smith}
\affil{Department of Physics and Astronomy, Michigan State University,
East Lansing, MI 48824}
\author{Karen Kinemuchi}
\affil{Dept. of Physics and Astronomy, University of Wyoming, Laramie, WY 82071}

\begin{abstract}

NSVS 5222076, a thirteenth magnitude star in the Northern Sky Variability
Survey, was identified by Oaster as
a possible new double-mode RR Lyrae star.  We confirm the double-mode
nature of NSVS 5222076, supplementing the survey data
 with new $V$ band photometry.  
NSVS 5222076 has a fundamental
mode period ($P_0$) of 0.4940 day and a first overtone period ($P_1$) of about 
0.3668 day, giving a period ratio of $P_1/P_0 = 0.743$. In most
double-mode RR Lyrae stars, the amplitude of the first overtone mode 
pulsation is greater than that of the fundamental mode pulsation.  That is
not true for this star.  Its fundamental mode light curve has an amplitude
twice as large as that of the first overtone mode, a ratio very
rarely seen even among the double-mode RR Lyrae that do have
relatively strong fundamental mode pulsation. Data from the
literature are used to discuss the location in the Petersen diagram
of double-mode RR Lyrae
stars having strong fundamental mode pulsation.  Such stars tend to occur 
toward the
short period end of the Petersen diagram, and NSVS 5222076 is no exception
to this rule.

\end{abstract}


\newpage

\section{Introduction}

The Northern Sky Variability Survey \citep{W04} holds photometry on 
about
14 million objects brighter than apparent magnitude $V$ = 15.5, which were
accessible to the ROTSE-I telescope from its location in New Mexico.
\citet{O05} surveyed a selection of the brighter RR Lyrae stars within
the Northern Sky Variability Survey (NSVS), identifying variables that 
might show either the Blazhko
Effect or double-mode behavior.  She identified NSVS 5222076 
($\alpha_{2000} = 15^{h}46^{m}26^{s}$; $\delta_{2000} = +44^{o}18'45''$) as a
possible new double-mode RR Lyrae.  Double-mode RR Lyrae stars, also
known as RRd or RR01 stars, exhibit simultaneous pulsation in the
fundamental and first overtone radial modes \citep{S95}.
 In this paper, we use the NSVS
data and additional $V$ band observations to confirm the double-mode
behavior of NSVS 5222076, and to show that it exhibits unusually strong
fundamental mode pulsation for such a star.  We also discuss the
location in the Petersen diagram of double-mode stars that, like 
NSVS 5222076, have
a fundamental mode amplitude as large or larger than the amplitude
of their first overtone mode pulsation.

\section{Photometry and Analysis}

\subsection{NSVS Data}
We analyzed two sets of photometry of NSVS 5222076.  First is the NSVS data for
5222076, obtained from the website at http://skydot.lanl.gov/nsvs/nsvs.php.  
The NSVS data include 218
observations taken between MJD 51274 and MJD 51559 with the ROTSE-I
telescope, with a typical photometric
uncertainty of $\pm$ 0.03 mag.  The NSVS observations are tied to the Tycho
$V$ system \citep{W04}, but, because the ROTSE-I observations are unfiltered,
RR Lyrae light curves based on NSVS data do not exactly 
correspond to those that would be observed through standard Johnson
$V$ filters \citep{K05}.  A period search on the NSVS data
using the Period04 program \citep{L05} found a primary period of 0.4940 day,
with an uncertainty of about $\pm$ 0.0001 day.
The light curve in the upper panel of Figure 1 shows the NSVS 
observations, folded with the
0.4940 day period after the NSVS MJDs have been converted to heliocentric
Julian dates.  The curve shows a scatter larger than expected from the 
photometric errors in the NSVS, a good indication that a single period does not
fully describe the NSVS observations. 

The 0.4940 day period and its first six harmonics were removed from the
NSVS data, and the Period04 program was used to search for periodicities
in the residuals.  A period of 0.3669 days was identified, with an
uncertainty of about $\pm$ 0.0002 days.  
A simultaneous fit was then made for the 0.4940 day period, the
0.3669 day period, and their six larger harmonics.  The resultant fundamental
mode and first overtone light curves are shown
in Figure 2.  The amplitude ratio for the fundamental and first overtone
modes, $A_0/A_1$, is approximately 2.  No significant evidence for
additional periods was found.

\subsection{V-Band Data}
Because of the relatively small number of NSVS observations, we
obtained new $V$ band photometry of NSVS 5222076 using the 60-cm telescope of
the Michigan State University campus observatory. Between JD 2453414
and JD 2453522, we obtained 1570 photometric observations on 16 nights
using an Apogee Ap47p CCD camera.
 Aperture photometry was
obtained differentially compared to star 1343-0291412, where the star
identification is that of the United States Naval Observatory NOMAD
catalog (http://www.nofs.navy.mil/nomad/), which merges data
from several prior astrometric and photometric catalogs \citep{Z04}.  
The NOMAD compilation gives approximate $V$ and $B$ magnitudes of
13.2 and 13.6 for 1343-0291412, making it comparable in magnitude, though
perhaps slightly redder than NSVS 5222076.  
NOMAD star 1344-0283988 was used as a check star.
For it, the NOMAD compilation lists magnitudes
of $V$ = 13.63 and $B$ = 14.20.  The Guide Star Catalog 2 (GSC2.2)
 numbers for the comparison
and check stars are N1321133346 and N132113394, respectively.
In Table 1 we give  heliocentric Julian
Dates and differential magnitudes in the sense 
$\Delta V$ = $V(5222076) - V(comp star)$
for the Michigan State University observations.
The accuracy of each of
the observations depends somewhat on the sky conditions at
the time of observation, but is typically $\pm$ 0.015 mag.

We analyzed the $V$ data in a fashion similar to that employed
for the NSVS observations.
We conducted a period search on the MSU photometry, again finding a
primary period of 0.4940 $\pm$ 0.0001 day (frequency $f_0$= 2.0242).  When the MSU observations
are folded with this period, it is again clear that a single
period does not by itself fully describe the light curve behavior (Figure 1,
lower panel).
We used Period04 to prewhiten the data to remove $f_0$
and its next six harmonics.  A search on the prewhitened data identified
a period of 0.3668 $\pm$ 0.0002 day (frequency $f_1$= 2.7265).  These periods were identified
as the fundamental and first overtone radial modes, respectively, and
are very close to the values obtained from the NSVS data.

To see whether there were any additional periods present in the data,
we used Period04 to simultaneously fit the $f_0$ and $f_1$
frequencies and their next six harmonics.  In a further period search
on this prewhitened data, we identified small contributions from
combinations of $f_0$ and $f_1$.

The final fit is shown in Table 2, where only components having an
amplitude greater than 0.005 mag are listed.
The deconvolved light curves for the 0.3668 day period and the 0.4940
day period are shown in Figure 3.

The resultant amplitude ratio, $A_0/A_1$, is again 2.0.  However, the
$V$ band data allow the light curve shape of each mode to be much more
clearly seen.  It is usually the
first overtone radial mode rather than the fundamental mode that
is strongest among double-mode RR Lyrae stars.  In such cases, the light 
curve shape of the fundamental
mode is more symmetric than is usually seen for stars that pulsate
solely in the fundamental mode (RRab stars).  Here, however, we see
the asymmetric light curve typical of RRab pulsators, with a rise to
maximum much steeper than the decline to minimum.  We also note in
Figure 3, the appearance of a bump before maximum light in 
the light curve of the fundamental mode.  Bumps before the rise to maximum 
are seen in many
RRab light curves, though often not with the prominence shown in Figure 3.
We also note that there seems to be some scatter in the deconvolved
light curves beyond that expected of observational error alone.  In the
case of the fundamental mode light curve this is particularly 
apparent just before
the beginning of the pre-rise bump. 

The pre-rise bump in the light curves of RRab stars is usually believed
to be a consequence of shock wave phenomena, though there has been
debate over the relative roles of heating by infalling gas
and of reflection of a compression wave at the boundary of the
core \citep{G88,C92,F97}.  The
bump is often more prominent in the light curves of RRab stars with
large amplitudes.  In the case of RR Lyrae, which has a light curve
modulated in a 40-day Blazhko cycle, the bump is stronger when the
amplitude of the light curve is larger \citep{W49,Ko05}.  It is tempting to
attribute the pre-maximum bump in the fundamental mode light curve of 
NSVS 5222076 to the same cause.  The interplay between the two pulsation
modes might be expected to change the amount of the shock, perhaps resulting
in the light curve scatter seen prior to the bump.  A problem with this
hypothesis is that the fundamental mode amplitude in NSVS 5222076 is
smaller than that usually seen in RRab stars that have strong pre-rise
bumps.

\section{The Petersen Diagram}
Double-mode RR Lyrae stars generally fall along a sequence in the Petersen
diagram \citep{P73}, in which the ratio of the periods is plotted against the
fundamental mode period (see, for example, Figure 6b of \citet{Sz04}).
In Figure 4 we have drawn two versions of the Petersen diagram.  In the
lower panel we have plotted double-mode RR Lyrae stars with published
ratios of the fundamental to first overtone amplitude,
$A0/A1$, that are smaller than
1.0.  Not all identifications of double-mode RR Lyrae in the literature
provide amplitude ratio information, but in Figure 4 we have been able to
include
data on double-mode stars in the Large Magellanic Cloud \citep{So03}, the
Sculptor dwarf spheroidal \citep{K01}, the Draco dwarf spheroidal \citep{Smi05},
M15 \citep{P95,N85}, the galactic field \citep{W05,G01,C00,C91, M03, J82}, and
the Sagittarius dwarf spheroidal \citep{Cs01}.  

In the upper panel of
Figure 4 we have plotted double-mode RR Lyrae that have strong fundamental
mode pulsation, with an amplitude $A0/A1$ greater than or equal to 1.0.
We see that, while the variables with weaker fundamental mode pulsation
span the entire range of periods, the stars with
stronger fundamental mode pulsation occur only at fundamental periods shorter
than 0.51~day, although even in that portion of the Petersen diagram
the strong fundamental mode pulsators are in the minority.  Although
the shorter period end of the Petersen diagram is more densely
populated by all double-mode RR Lyrae stars in Figure 4, the 
concentration of the strong fundamental mode pulsators to that
region appears to
be too strong to be a statistical fluke.  Table 3 shows the number
and percentage
of double-mode RR Lyrae within this diagram that have $A0/A1$ greater
than or equal to 1.0, dividing the sample into three subgroups
based on fundamental mode period. Stars with
$A0/A1$ greater than or equal to 1.0 account for 12 $\pm$ 2 percent of the
complete sample, but the percentage rises at shorter periods and falls
at longer periods. A chi squared test for equality of distribution
indicates a less than 1 percent probability of obtaining the observed
excess of strong fundamental mode pulsators at periods shorter
than 0.47 day or the deficiency of such stars at periods longer than 0.51
day.

Of the 344 double-mode RR Lyrae stars used to construct figure 4, only seven
have reported ratios $A0/A1$ greater than 1.5.  Only three have a ratio
greater than 1.8, making them comparable to NSVS 5222076. These three
stars are all among the sample of double-mode RR Lyrae in the Large Magellanic
Cloud. Not plotted in Figure 4 because of the large overlap with the
\citet{So03} dataset, the LMC double-mode stars identified by \citet{A97}
also indicate that double-mode stars with a ratio $A0/A1$ approaching
2 are very rare.

One caveat ought to be noted.  A large fraction of the RR Lyrae stars
in this sample, particularly at periods shorter than 0.51 day, are
located in the Large Magellanic Cloud. It is possible that the properties of
the double-mode RR Lyrae stars are closely tied to the
particular horizontal branch properties of their parent system.
If that is the case, the correlations reported here may not be universal. 

One might also wonder whether any of the field RR Lyrae stars of the Galaxy
in Figure 4 might actually be double-mode Cepheids rather than double-mode
RR Lyrae stars. As shown, for example, in Figure 1 of \citet{So0}, 
double-mode Cepheids in the Magellanic Clouds can have period ratios similar to the
0.74 seen among the RR Lyrae stars.  However, in such cases, the
fundamental mode periods are longer than 0.7 days, so that if there were
Galactic Cepheids similar to those in the Magellanic Clouds, they
would be expected to fall off the long period end of Figure 4.

NSVS 5222076 is plotted as a cross in figure 4.  It falls slightly to the
small period ratio side of the general Petersen diagram trend, but
is not greatly discrepant. We note the presence of three more
discrepant stars in figure 4, all with reportedly strong fundamental
mode components.  The dark point in the upper panel at a ratio
of 0.7493 represents one of the stars in the LMC sample.  We have
also plotted in figure 4 the locations of two double-mode RR Lyrae
in M3, V13 and V200, which have unusually small reported period ratios
\citep{Cl04}.
One complication in the cases of these two stars is that
\citet{Cl04} find evidence that the amplitude ratios of the two
stars may change with time (though note the caution in \citet{B05}).
The issue of the extent to which the relative amplitudes of the two
components in double-mode RR Lyrae stars can change over time remains
to be established.

\begin{acknowledgements}

We thank the National Science Foundation for partial support of this work
under grants AST-0205813, AST-0440061, and AST-0307778.  We thank the
referee for several very useful comments.

\end{acknowledgements}

\clearpage
\begin{figure}
\plotone{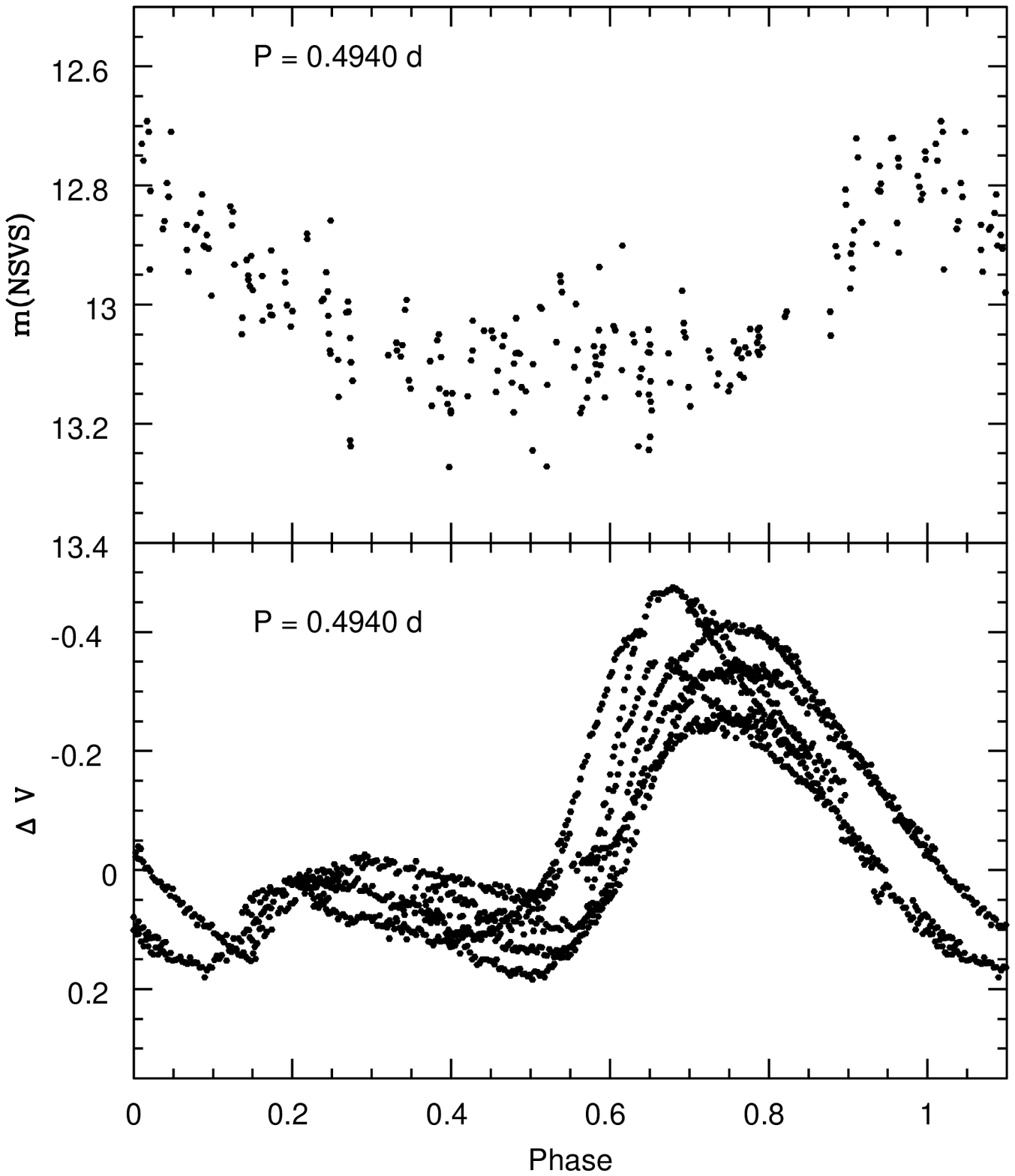}
\figcaption[fig1.eps]{The upper panel shows the light curve of 
NSVS 5222076 based on observations in the Northern
Sky Variability Survey, phased with a period of 0.4940~d.  The lower
panel shows the MSU $V$ band data, phased with the same period.
\label{fig1}}  
\end{figure}

\begin{figure}
\plotone{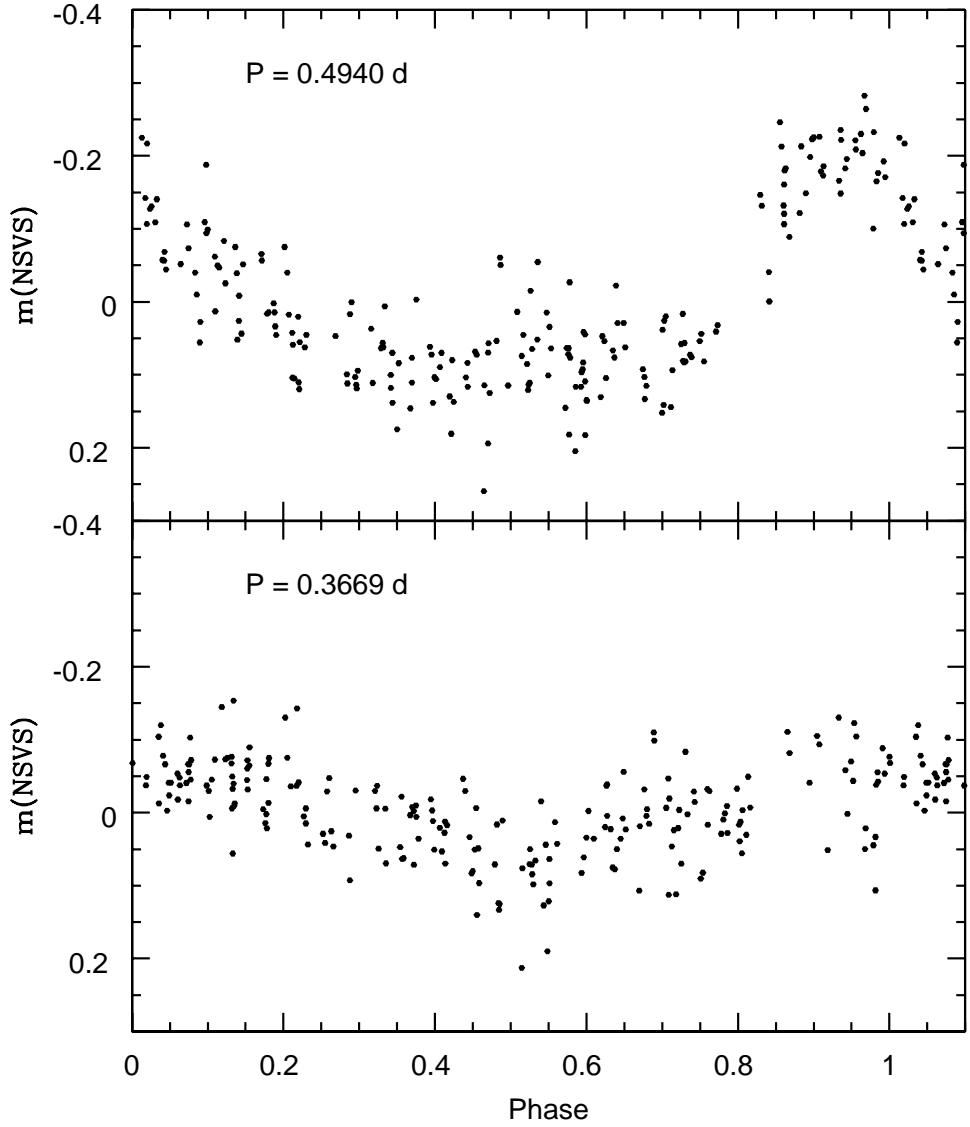}
\figcaption[fig2.eps]{Fundamental mode and first overtone mode light
curves of NSVS 5222076, based upon NSVS data, and phased with the indicated
periods.
 \label{fig2}}
 \end{figure}
 
\begin{figure}
\plotone{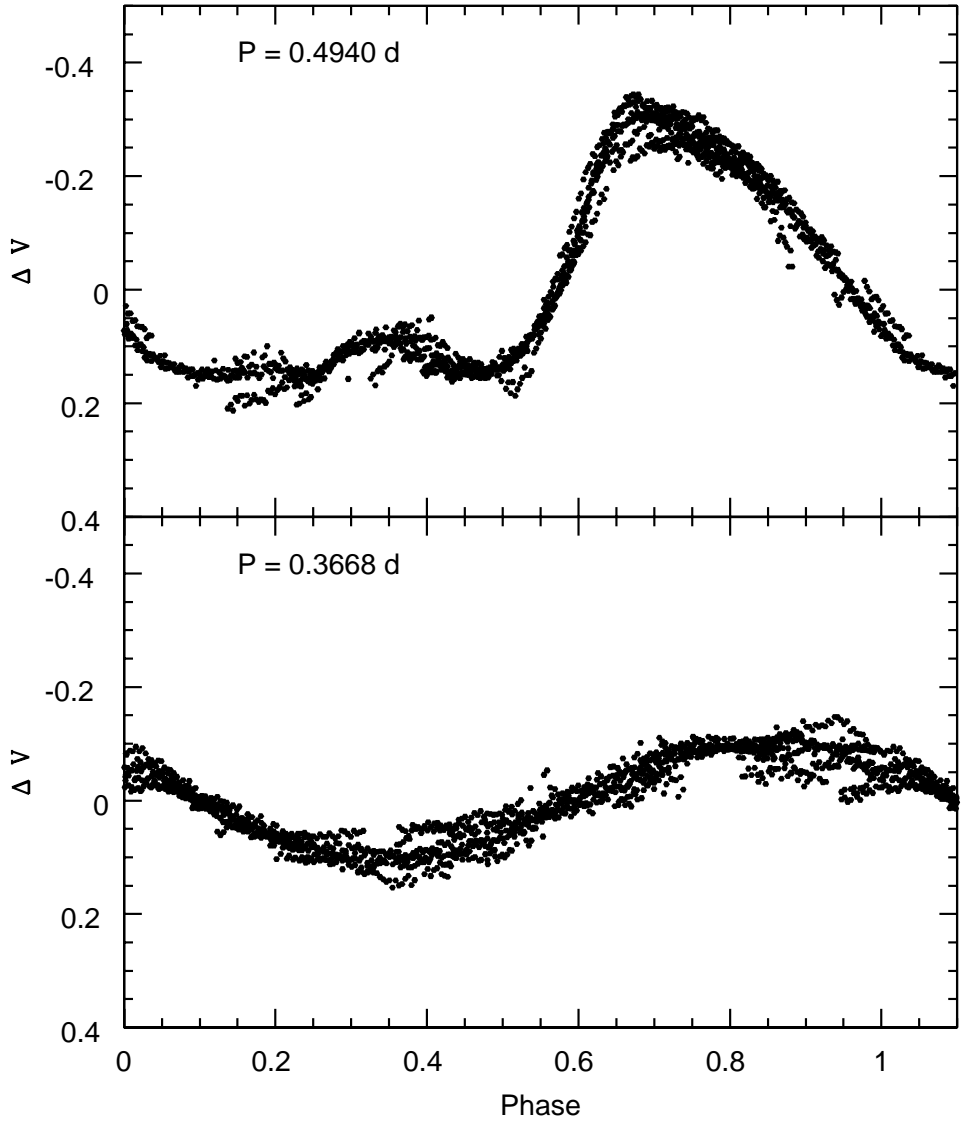}
\figcaption[fig3.eps]{Fundamental mode and first overtone mode light
curves of NSVS 5222076, based on Michigan State University data.
\label{fig3}}
\end{figure}

\begin{figure}
\plotone{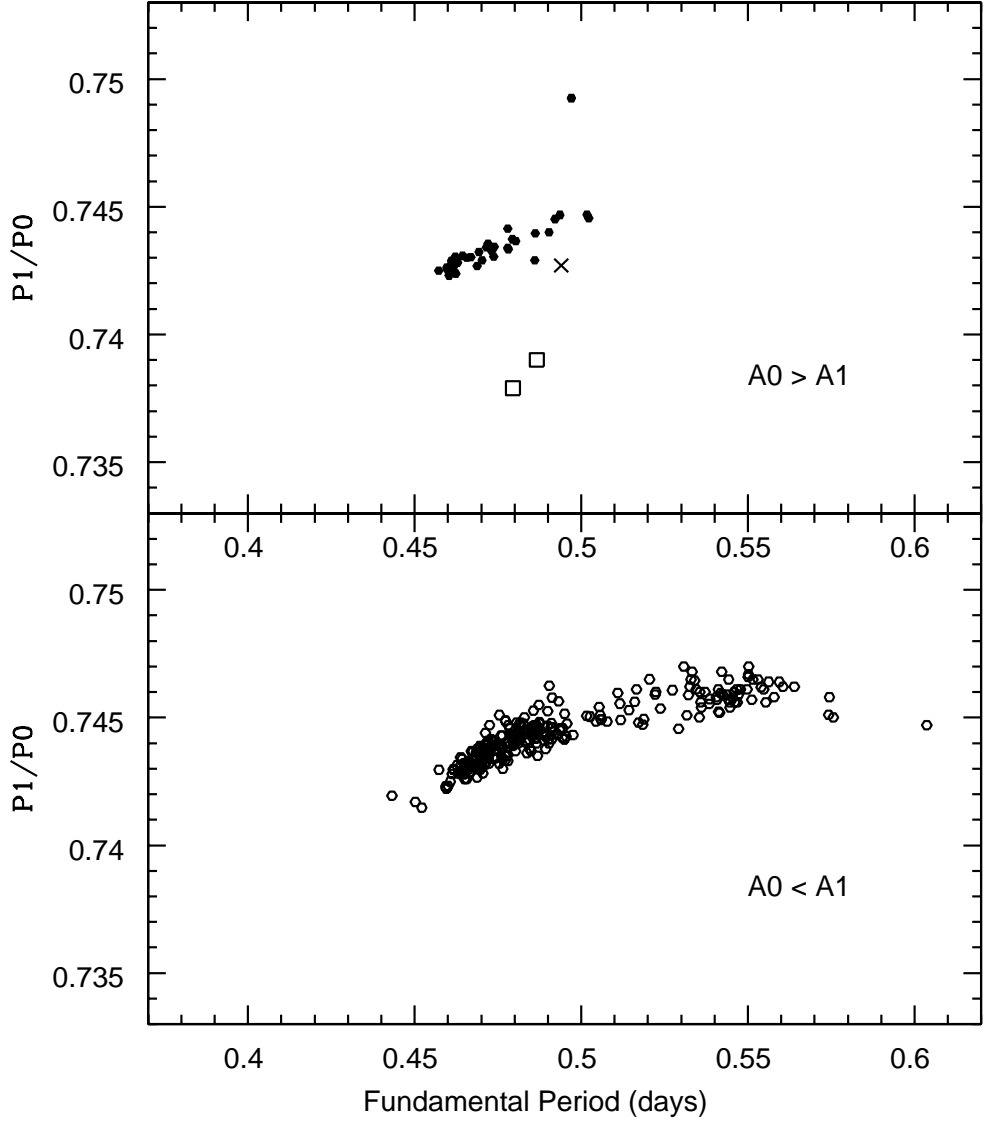}
\figcaption[fig4.eps]{Petersen diagrams for a sample of double mode
RR Lyrae from the literature for which amplitude information
is available.  The upper panel depicts double-mode stars for which the
fundamental mode amplitude, $A0$, is stronger than the first overtone
mode amplitude, $A1$.  Variables with weaker fundamental mode amplitudes
are plotted in the lower panel.  In the upper panel, X marks the location
of NSVS 5222076 and the two open squares indicate the places of two
unusual double-mode RR Lyrae in the globular cluster M3.
\label{fig4}}
\end{figure}

\clearpage
\begin{deluxetable}{cc} 
\tablewidth{0pc}
\footnotesize
\tablenum{1}
\tablecaption{CCD Photometry (Michigan State University)}
\tablehead{
\colhead{HJD} & \colhead{$\Delta$V}
          }
\startdata
2453414.8263  &  -0.233\\
2453414.8275  & -0.251 \\
2453414.8287  &  -0.255\\
2453414.8299  & -0.243\\
\enddata
\tablecomments{The complete version of this table is in the electronic 
edition of the Publications.  The printed edition contains only a sample.}
\end{deluxetable}

\clearpage

\clearpage
\begin{deluxetable}{cl} 
\tablewidth{0pc}
\footnotesize
\tablenum{2}
\tablecaption{Fit to $V$ Band Data}
\tablehead{
 \colhead{Frequency (c/d)} & \colhead{Ampl. (mag)}
          }
\startdata
2.0242 ($=f_0$) & 0.203\\
$2f_0$ & 0.088\\
$3f_0$ & 0.045\\
$4f_0$ & 0.019\\
$5f_0$ & 0.014\\
$6f_0$ & 0.008\\
$7f_0$ & 0.005\\
2.7265 (=$f_1$) & 0.105\\
$2f_1$ &  0.010\\
$3f_1$ & 0.006\\
$f_0 + f_1$ & 0.021\\
$2f_0 + f_1$ & 0.036\\
$3f_0 + f_1$ & 0.018\\
$f_0 - f_1$ & 0.024\\

\enddata
\end{deluxetable}

\begin{deluxetable}{crr} 
\tablewidth{0pc}
\footnotesize
\tablenum{3}
\tablecaption{RRd Stars with $A0/A1$ $\geq$ 1.0}
\tablehead{
\colhead{Fundamental Period (days)} & \colhead{Number} & \colhead{Percentage}
          }
\startdata
$< 0.47$ & 21 & 24 \\
0.47 -- 0.51 & 20 & 11 \\
$> 0.51$  & 0 & 0\\
All periods & 41 & 12 \\
\enddata
\end{deluxetable}

\clearpage

\end{document}